\documentclass[12pt]{iopart}
\usepackage{graphicx}
\usepackage{bigstrut}

\expandafter\let\csname equation*\endcsname\relax
\expandafter\let\csname endequation*\endcsname\relax

\usepackage{amsmath}

\newcommand{\beq}{\begin{equation}}
\newcommand{\eeq}{\end{equation}}
\def\bsp#1\esp{\begin{split}#1\end{split}}
\newcommand{\nlo}      {{\rm NLO}}
\newcommand{\lhe}      {{\rm LHE}}
\newcommand{\helacnlo}{\texttt{HELAC-NLO}}
\newcommand{\helaconeloop}{\texttt{HELAC-1loop}}
\newcommand{\helaconeloopdd}{\texttt{HELAC-1loop@dd}}

\newcommand{\cuttools}{\texttt{CutTools}}

\newcommand{\qd}{\texttt{QD}}
\newcommand{\powhegbox}{\texttt{POWHEG-Box}}
\newcommand{\powhel}{\texttt{PowHel}}

\newcommand{\gev}{\ensuremath{\,\mathrm{GeV}}}

\newcommand{\ud}{\mathrm{d}}

\newcommand{\ttbb}{t$\bar{{\rm t}}$\,b$\bar{{\rm b}}$}
\newcommand{\muf}   {\ensuremath{\mu_{\rm F}}}
\newcommand{\mur}   {\ensuremath{\mu_{\rm R}}}
\newcommand{\rad}      {\mathrm{rad}}
\newcommand{\mbb}   {\ensuremath{m_{\rm b\,\bar{\rm b}}}}
\newcommand{\pT}{\ensuremath{p_{\perp}}}

\newcommand{\pTb}{\ensuremath{p_{\perp,{\rm b}}}}

\newcommand{\pTbb}{\ensuremath{p_{\perp,\bar{\rm b}}}}

\newcommand{\pTmin}[1]{\ensuremath{p_{\perp #1}^{\min}}}

\newcommand{\kT}{\ensuremath{k_{\perp}}}

\newcommand{\mt}{\ensuremath{m_{\rm t}}}

\newcommand{\rt}{{\rm t}}
\newcommand{\bt}{\ensuremath{\bar{{\rm t}}}}
\newcommand{\rb}{{\rm b}}
\newcommand{\bb}{\ensuremath{\bar{{\rm b}}}}
\newcommand{\bq}{\ensuremath{\bar{{\rm q}}}}

\newcommand{\tB}{\ensuremath{\widetilde{B}}}
\newcommand\Ref[1]     {Ref.\,\cite{#1}}

\newcommand\eqn[1]     {(\ref{#1})}
\newcommand\fig[1]     {Figure~{\ref{#1}}}
\newcommand\figs[2]    {Figures~{\ref{#1}} and ~\ref{#2}}

\begin{document}


\title{Hadroproduction of t anti-t pair with a b anti-b pair using PowHel}

\author{A.~Kardos$^1$ and Z.~Tr\'ocs\'anyi$^2$}
\address{
  $^1$ INFN, Sezione di Milano-Bicocca, I 20126 Milano, Italy}
\address{
  $^2$ Institute of Physics and MTA-DE Particle Physics Research Group,\\
  University of Debrecen, H-4010 Debrecen P.O.Box 105, Hungary}

\ead{Adam.Kardos@mib.infn.it, Z.Trocsanyi@atomki.mta.hu}

\begin{abstract}
We simulate the hadroproduction of a t\bt\ pair in association with a
b\bb\ pair at 14\,TeV LHC using the \powhel\ package.  We use the
generated events, stored according to the Les-Houches event format, to
make predictions for differential distributions formally at the
next-to-leading order (NLO) accuracy and we compare these
to existing predictions accurate at NLO.
\end{abstract}

\pacs{12.38.-t, 14.65.Ha, 14.65.Fy}
\submitto{\JPG}

\maketitle

\section{\label{sec:introduction} Introduction}

According to the latest, most precise measurements, the t-quark mass
is $m_{\rt} = 173.5\pm0.6\pm0.8$\,GeV \cite{Beringer:1900zz},
indicating that the Yukawa coupling of the t-quark, $y_{\rt}=0.997\pm0.008$
equals one with better than 1\,\% accuracy.  This remarkable result
suggests that it is important to measure this coupling also directly as
precisely as possible. After the recent discovery of a Higgs
boson at the LHC \cite{ATLAS:2012ae,Chatrchyan:2012tx}, the focus of
analyses is shifting towards determining whether the couplings of this
new particle to other fundamental particles agree with the Standard
Model (SM) expectations. The measurement of the t\bt$H$ coupling is
important also because deviation from the SM expectation provides
a signal of physics beyond the SM. The Higgs boson with mass
approximately 125\,GeV does not decay into a pair of t-quarks,
therefore the only process to measure the t\bt$H$ coupling in
a model independent way is the study of hadroproduction of the
Higgs-boson in association with a t\bt\ pair
\cite{Maltoni:2002jr,Belyaev:2002ua}.

The produced Higgs-boson decays immediately and only its decay products
can be observed.  According to the SM predictions, such a Higgs-boson
decays dominantly into a b\bb\ pair. Thus one can plan to measure
the t\bt$H$ coupling in the pp $\to \rt\bt H \to \rt\bt\,\rb\bb$
process. Unfortunately, this process has a large background from the
direct QCD process pp $\to \rt\bt\,\rb\bb$. In order to make a proper
background estimation and optimize the experimental selection of the
t\bt$H$ events, one needs simulation of the QCD background with high
precision.

The most up to date method to simulate LHC processes with high precision
is matching NLO QCD predictions to shower Monte Carlo programs (SMC).
Presently, two formalisms are used frequently to match matrix
element (ME) calculations at the NLO accuracy to parton showers (PS),
the MC@NLO~\cite{Frixione:2002ik, Frixione:2010wd} and POWHEG
\cite{Nason:2004rx, Frixione:2007vw} methods. Indeed, both methods were
used to make predictions for t\bt$H$ hadroproduction
\cite{Frederix:2011zi,Garzelli:2011vp}, and the two predictions were
found to be in agreement in \Ref{Dittmaier:2012vm}. Here we make one
step further and describe our implementation of the pp $\to$ \ttbb\
process within the \powhel\ framework.
 
The \powhel\ framework combines the \powhegbox~\cite{Alioli:2010xd}, a
flexible program implementing the POWHEG method, and the
\helacnlo~package \cite{Bevilacqua:2011xh}. The output of \powhel\ is
simulated events stored according to the Les Houches accord
\cite{Alwall:2006yp} (LHE).  Those events can be fed into any shower
Monte Carlo (SMC) program for generating events with hadrons.  Using
the same framework we already provided LHE's for several processes at
the LHC, namely to the production of a t\bt-pair in association with a
hard object, such as a jet \cite{Kardos:2011qa}, a scalar boson
\cite{Garzelli:2011vp}, or a vector boson \cite{Garzelli:2012bn}.

The scope of this article is to present the simulation of the \ttbb\ LHE's.
The corresponding predictions at the NLO accuracy were first presented in
Refs.~\cite{Bredenstein:2009aj,Bevilacqua:2009zn,Bredenstein:2010rs}.
The process under consideration represents an example of
unprecedented complexity in \powhel, which raises some
technical difficulties, worth of detailed description. 

\section{\label{sec:method} Method}

We use the POWHEG formula (see Eq.~(3.4) in \Ref{Frixione:2007vw}) to
estimate the cross section from events with either unresolved or
resolved first radiation,
\beq
\bsp
\ud \sigma_{\lhe} = \tB(\Phi_B)\ud \Phi_B
\Bigg[&\Delta(\Phi_B, \pTmin{})
+ \ud \Phi_{\rad}
\Delta\Big(\Phi_B, \kT(\Phi_R)\Big)
\\& \times
\frac{R(\Phi_R)}{B(\Phi_B)}
\Theta(\kT(\Phi_R)-\pTmin{})\Bigg]
\,.
\label{eq:sigmaLHE}
\esp
\eeq
In \eqn{eq:sigmaLHE} $\tB(\Phi_B)$ denotes the NLO-corrected fully
differential cross section belonging to the underlying Born
configuration $\Phi_B$  (the integration over the momentum fractions is 
included implicitly),
\beq
\tB(\Phi_B) =
B(\Phi_B) + \mathcal{V}(\Phi_B) + \int\!\ud\Phi_{\rad}\mathcal{R}(\Phi_R)
+ \int\!\frac{\ud x}{x}\Big[G_{\oplus}(\Phi_B)+G_{\ominus}(\Phi_B)\Big]
\,,
\label{eq:tB}
\eeq
and $\Delta(\Phi_B, \pT)$ is the POWHEG Sudakov form factor that
exponentiates the integral of the ratio of the real radiation
$R(\Phi_R)$ and Born $B(\Phi_B)$ contributions over the radiation phase
space,
\beq
\Delta(\Phi_B, \pT) = \exp
\left\{-\int\!\ud \Phi_{\rad}\frac{R(\Phi_R)}{B(\Phi_B)}
\Theta(\kT(\Phi_R)-\pT)
\right\}
\,.
\label{eq:sudakov}
\eeq
In \eqn{eq:tB} $\mathcal{R}$ denotes the difference of the real
emission part and the subtraction terms, 
$\mathcal{R}(\Phi_R) = R(\Phi_{n+1}) - C(\Phi_{n+1})$. Similarly,
$\mathcal{V}$ denotes the regularized (finite in $d=4$ dimensions)
virtual correction (was $V(\Phi_n)$ in Eq.~(3.2) in
\Ref{Frixione:2007vw}), i.e., it also contains the integrated
subtraction terms, while we use $V(\Phi_n)$ to denote the unregularized
virtual correction.  Using $\tB(\Phi_B)$, we obtain the cross section at
NLO accuracy as an integral over the Born phase space,
\beq
\sigma_{\nlo} = \int\!\ud\Phi_B\tB(\Phi_B)
\,,
\label{eq:sigmaNLO}
\eeq

We used the \helacnlo~package to generate the crossing symmetric matrix
elements required in the \powhegbox\ as input. In particular, (i) the
squared matrix elements for the flavour structures of the Born
(gg$\to$ t\bt\:b\bb, q\bq $\to$ t\bt\:b\bb, \bq q$\to$ t\bt\:b\bb) and
real radiation emission (q\bq$\to$ t\bt\:b\bb g, gg$\to$ t\bt\:b\bb g,
\bq g$\to$ t\bt\:b\bb\bq, g\bq$\to$ t\bt\:b\bb\bq, \bq q$\to$
t\bt\:b\bb g, qg$\to$ t\bt\:b\bb q, gq$\to$ t\bt\:b\bb q)
subprocesses (q$\in$\{u,d,c,s\} -- we neglect the contribution of
b-quarks in the initial state), (ii) the colour-correlated and
spin-correlated squared matrix elements for the Born flavour
structures, and (iii) the finite part of the virtual correction
contributions in dimensional regularization, on the basis of the OPP
method \cite{Ossola:2006us} complemented by Feynman rules for the
computation of the QCD $R_2$ rational terms~\cite{Draggiotis:2009yb}.
With this input \powhegbox\ generates events with either unresolved or
resolved first radiation. Then, one can use any shower Monte Carlo
(SMC) program for generating events with hadrons.  

\subsection{\label{sec:checks} Predictions at NLO accuracy}

The process presented here was studied at the NLO accuracy in
the literature~\cite{Bevilacqua:2009zn,Bredenstein:2010rs}, which
enabled us to make detailed checks of our calculation. We reproduced
all figures of \Ref{Bevilacqua:2009zn} and found complete agreement.

As this process has different hard scales in the final state the proper
choice of the scale is important as the convergence of the perturbation
series is influenced by that. For instance, with the central scale
$\mur = \muf = \mu_0 = \mt$, the NLO correction is large, about 80\,\%
\cite{Bevilacqua:2009zn}. It was pointed out in \Ref{Bredenstein:2010rs}
that the convergence is much better with the central scale
$\mu_0^2 = \mt \sqrt{p_{\bot,b}p_{\bot,\bar{b}}}$, when NLO
corrections amount to about 24\,\%. This scale was inspired by scale
choices made in shower Monte Carlo programs and is designed to mimic
the effect of higher order corrections in the low-order prediction. It
is close to the scale where the scale dependence at NLO accuracy
reaches its maximum, leaving little room for positive higher order
corrections: the predictions with the central scale appear near the
higher edge of the scale-dependence bands.

\begin{figure*}[t!]
\begin{center}
\includegraphics[width=0.7\linewidth]{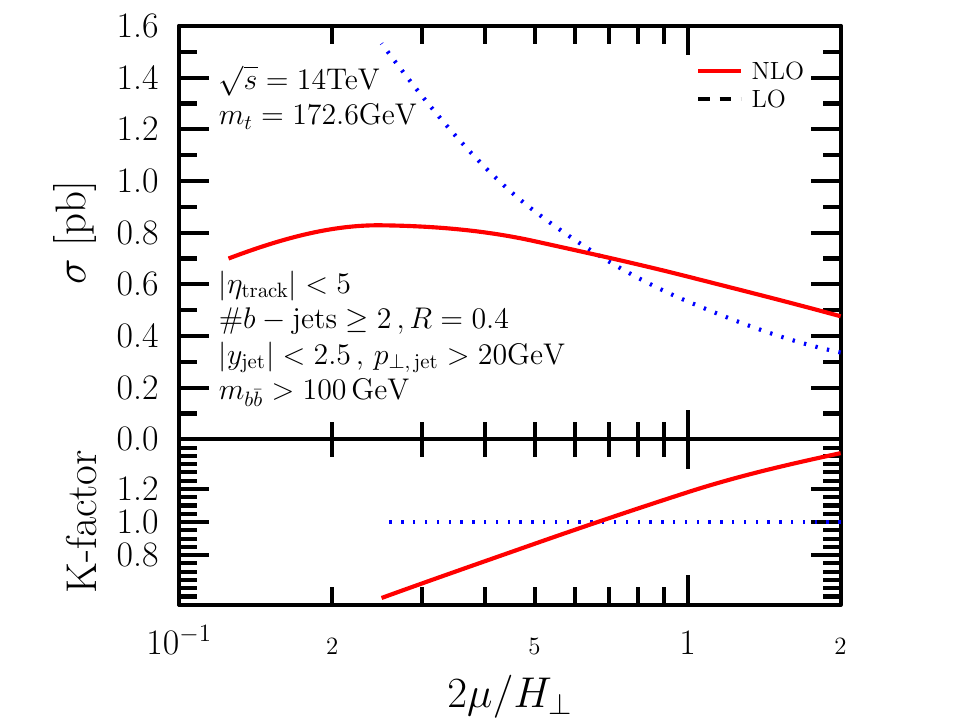}
\end{center}
\caption{Dependence of the total cross section at LO and NLO accuracy,
with cuts shown in the legend.
}
\label{fig:sigmatot}
\end{figure*}

As our goal is to generate the effect of higher-order corrections by
the shower, in generating pre-showered events ready for feeding those
into SMC's, we find more appropriate to use the central scale
$\mu_0 = H_\bot/2$, where $H_\bot$ is the sum of transverse masses
$m_{\bot,f} = \sqrt{m_f^2+p_{\bot,f}^2}$ of the final-state partons in
the hard-scattering event, $H_\bot  = \sum_f m_{\bot,f}$ (the sum
includes four partons at LO and in the virtual correction, while five
partons in the real correction). This choice reproduces better the
standard choice for the t\bt\ final state near threshold production of
the final state: as $p_{\bot,{\rm b}}$ and $p_{\bot,\bar{\rm b}} \to 0$, 
$\mu_0 \to m_{\bot,{\rm t}}$.  In \fig{fig:sigmatot} we show our
predictions for the total cross section with the following selection
cuts (setup I in \Ref{Bredenstein:2010rs}):
\begin{enumerate}
\item A massless parton was considered as a possible jet constituent if
its pseudorapidity is within the range [-5,+5] ($|\eta|<5$).
\item Jets were reconstructed with the \kT-algorithm using
$\pTmin{,{\rm jet}} = 20$\,GeV and $R=0.4$.
\item We required at least one $b$- and one
$\bar{b}$-jet, with $|y_{b\,(\bar{b})}|<2.5$. (We distinguish between
quark flavours and tagged jets using roman typsetting for the former
and italic for the latter.)
\item Events with invariant mass of the $b\bar{b}$-jet pair below
$m^{\min}_{b\,\bar{b}} = 100\,\gev$ were discarded.
\end{enumerate}

\begin{figure*}[t!]
\begin{center}
\includegraphics[width=0.49\linewidth]{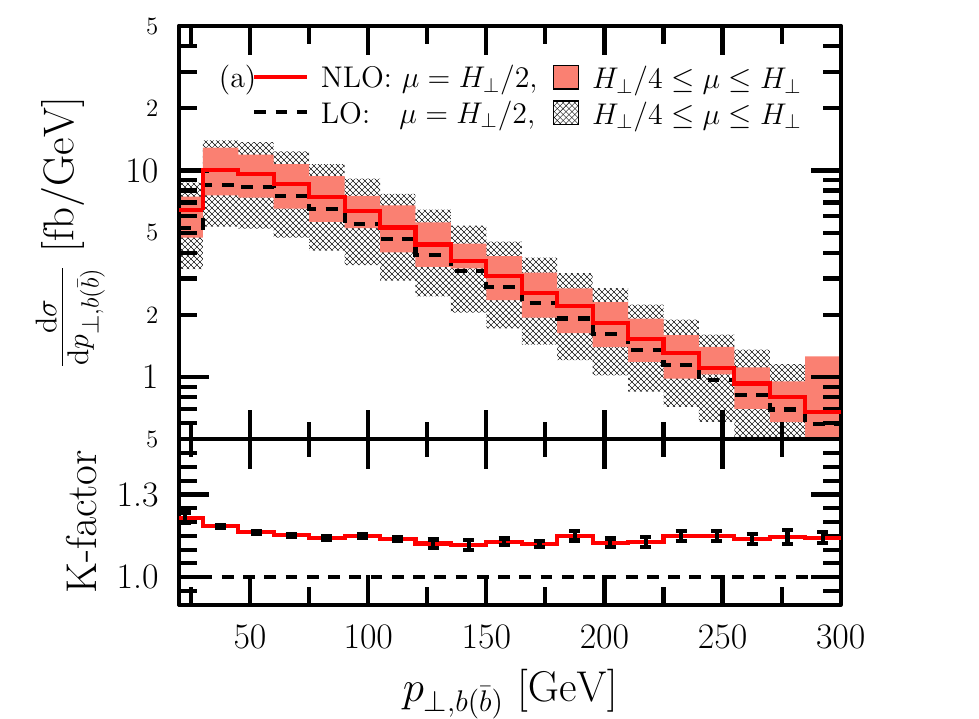}
\hfill
\includegraphics[width=0.49\linewidth]{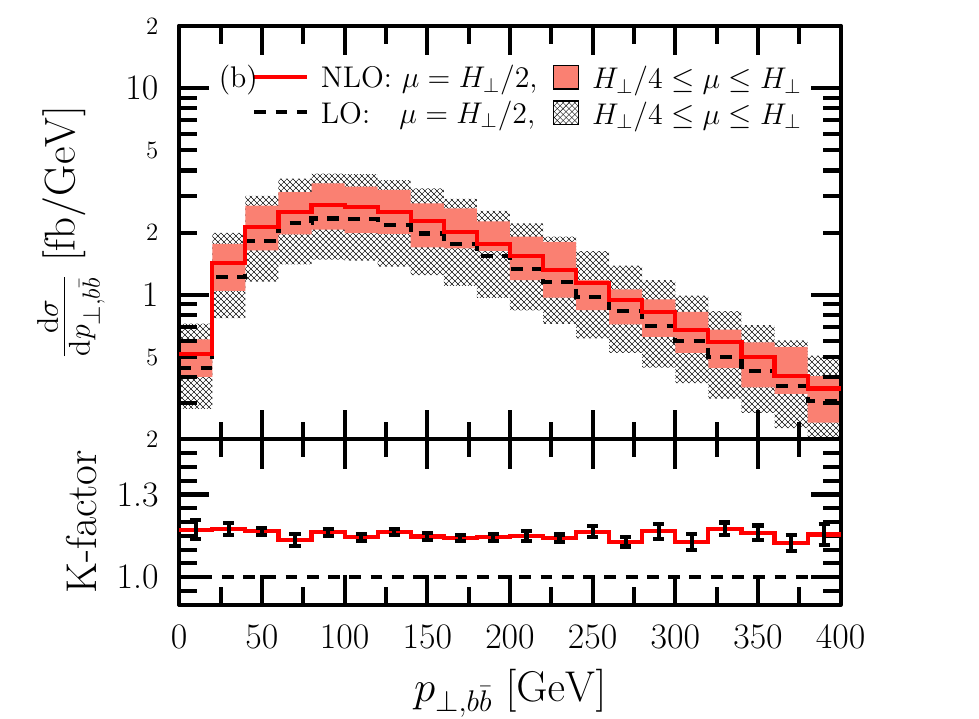}
\end{center}
\caption{(a) Transverse momentum distribution of the $b$-jet and (b)
that of the $b\bar{b}$-jet pair at the LHC at $\sqrt{s} = 14$\,TeV using
\powhel. The shaded bands correspond to cross sections obtained with
varying the scale around the default one in the range $[\mu_0/2,2 \mu_0]$.  
The lower panels show the NLO predictions normalized by the predictions
at LO, with errorbars representing the statistical accuracy of the
numerical integration.
}
\label{fig:ptb-nlo}
\end{figure*}

Clearly, our choice for the default scale gives as stable predictions at
NLO as those with $\mu_0^2 = \mt \sqrt{p_{\bot,b}p_{\bot,\bar{b}}}$.
The cross section with this default scale is 534\,fb at LO and 630\,fb
at NLO, so the NLO K-factor is 1.18, and the NLO scale-uncertainty in
the range $[\mu_0/2, 2\mu_0]$ is $^{+22\,\%}_{-32\,\%}$.  The shapes of
the distributions are very similar at LO and NLO accuracies as seen in
\figs{fig:ptb-nlo}{fig:mbb-nlo}. In the same figures we also show the
bands representing the scale dependence due to the variation of the
scale in the range $[\mu_0/2, 2\mu_0]$. We find that the histograms
with the default scale run in the middle of the band representing the
scale dependence.
\begin{figure*}[t!]
\begin{center}
\includegraphics[width=0.49\linewidth]{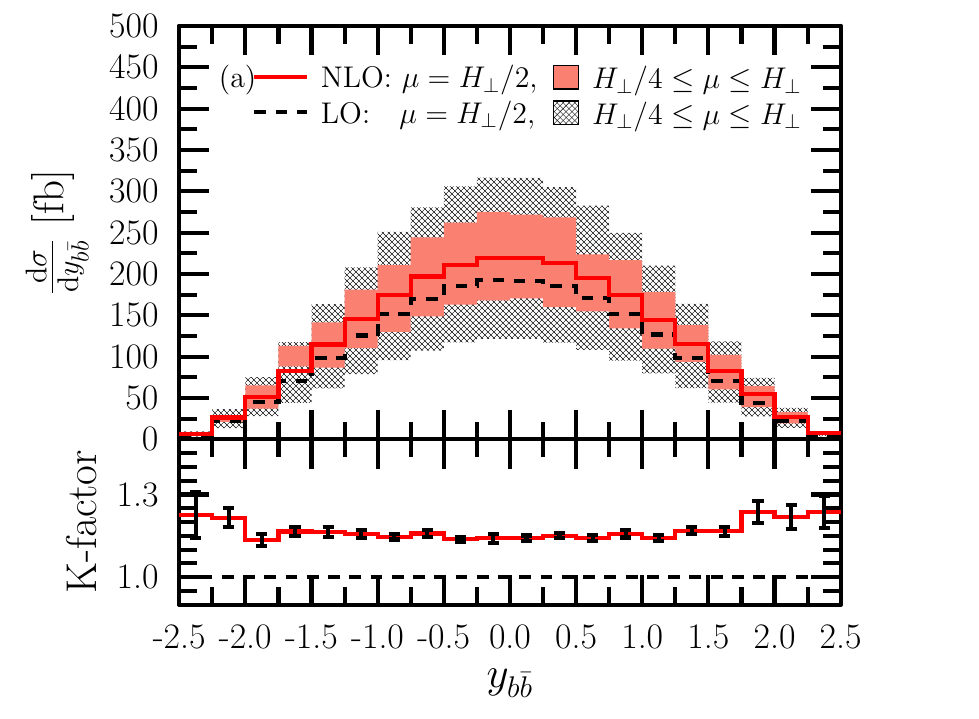}
\hfill
\includegraphics[width=0.49\linewidth]{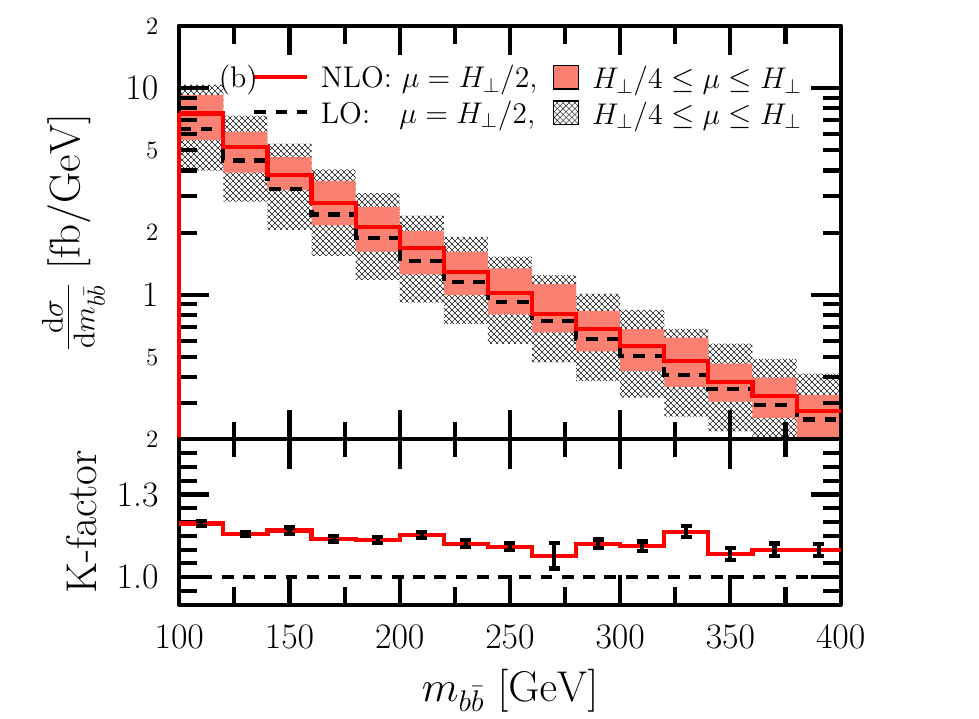}
\end{center}
\caption{Same as \fig{fig:ptb-nlo} for (a) the rapidity distribution and
(b) invariant mass distribution of the $b\bar{b}$-jet pair.
}
\label{fig:mbb-nlo}
\end{figure*}

\subsection{Generation cuts and suppression}

As we treat the b-quarks massless, the fully differential Born cross
section becomes singular if any of the transverse momenta of the
b-quarks $\pTb$ or $\pTbb$, or the invariant mass of the b\bb-pair
$\mbb$ vanishes.  To make the integral of $B$ finite over the whole
Born phase space, we introduce generation cuts by requiring $\pTb \geq
\pT^{(\mathrm{g.c.})}$ and $\mbb \geq \mbb^{(\mathrm{g.c.})}$, where
$\pT^{(\mathrm{g.c.})} = \mbb^{(\mathrm{g.c.})} =  2\gev$. We checked
that these generation cuts are sufficiently low, so that the physical
predictions are independent of those when selection cuts characteristic
to physical analyses are superimposed. However, with such low generation
cuts the event generation is very inefficient because most of the events
are generated in regions of the phase space where $B$ is
large, but those regions are usually not selected in the analyses.
Therefore, we also introduce a suppression factor \cite{Alioli:2010qp} for
$\tB$ of the form
\begin{equation}
\mathcal{F}_{\mathrm{supp}} =
\left(\frac{\mbb^2}{\mbb^2 + (\mbb^{\mathrm{supp}})^2}\right)^3
\prod_{i={\rm b},\bar{{\rm b}}}
\left(\frac{p_{\perp,i}^2}{p_{\perp,i}^2 + (\pT^{\mathrm{supp}})^2}\right)^3
\,,
\label{eq:suppression}
\end{equation}
where $\pT^{(\mathrm{supp})} = \mbb^{(\mathrm{supp})} = 30\gev$. Such a
suppression factor does not influence physical observables, which we
checked explicitly.

\subsection{Precision and efficient evaluation of loop amplitudes}

For this process there is the high degree of difficulty
of evaluating the loop amplitudes. The high-rank of the tensor
loop-integrals makes the numerical computation of the loop amplitude
unstable, and thus, unreliable when double precision arithmetics is
used in \cuttools\ \cite{Ossola:2007ax}, as implemented in
\helaconeloop\ (part of \helacnlo).  In order to control numerical
instabilities we employ an $\cal N = \cal N$ test as implemented in
\cuttools. This means that for a given numerator we determine the
scalar-integral coefficients using double precision arithmetics. We
check the accuracy of the integrand by reconstructing it using all the
coefficients (and spurious terms) with a randomly chosen loop momentum.
If the reached relative accuracy is worse than $10^{-4}$, we pass the
same phase space point in double-double precision (computed
simultaneously with the double-precision version) to \helaconeloopdd\
to recalculate all the coefficients (using the multiple precision part
of \cuttools). The \helaconeloopdd\ code is a straightforward extension
of \helaconeloop\ to double-double precision using the package \qd\
\cite{hida00}. This way we solved all numerical instabilities in the
computation of the virtual correction.  
The above procedure however, makes the numerical computation of the loop
amplitudes rather cumbersome, requiring much more CPU time. This problem is
magnified by the hit-and-miss procedure, the way event generation is
performed in the \powhegbox. During NLO integration the inclusive
NLO cross section is computed simultaneously with the maximal value of \tB
\beq
\tB_{\max} = \max_{\Phi_B}\tB(\Phi_B)
\,.
\eeq
When the underlying Born kinematics $\Phi_B$ is generated for an event,
it is accepted if $\tB(\Phi_B)\ge\xi\tB_{\max}$, where $\xi$ is a
random number picked between zero and one.
Using this hit-and-miss method, the \tB\ function has to be evaluated
several times to find a suitable phase space point which is selected.
If \tB\ is calculated multiple times per event, event generation
becomes highly inefficient because \tB\ contains the virtual part,
whose evaluation is very time consuming.  To improve the efficiency, we
first generate events with a fake virtual contribution, proportional to
the Born cross section, $V = \alpha B$. The value of $\alpha$ is
arbitrary. We select those events that pass the selection cuts, whose
weights in a final step are reweighted by the true virtual contribution.
In practice, we choose $\alpha$ such that the physical cross section
changes only little with this reweighting. We have not made a
systematic study to minimize the effect of reweighting, but chose
$\alpha = 1/2$ after some simple test runs. Using the fake virtual, we
can reduce the computation of the virtual contributions to few hundred
thousand phase space points. The plots shown in the next section are
based upon about 100\,k LHE's.

The POWHEG method produces unweighted events. When we generate events
with the fake virtual part, unweighting must be done using the NLO cross
section obtained also with the fake virtual. After the events that pass
the selection cuts are collected, we reweight those 
\beq
\mathcal{W}\big|_{V = B/2} \to \mathcal{W} = \mathcal{W}\big|_{V = B/2}
\frac{\tB(\Phi_B)}{\tB\big|_{V = B/2}(\Phi_B)}
\,,
\eeq
where $\mathcal{W}\big|_{V = B/2}$ is the weight with fake virtual, and
$\mathcal{W}$ is the weight with the true virtual part.  Clearly, the
new weight becomes dependent upon the underlying Born kinematics.

\begin{figure}[t!]
\begin{center}
\includegraphics[width=0.7\linewidth]{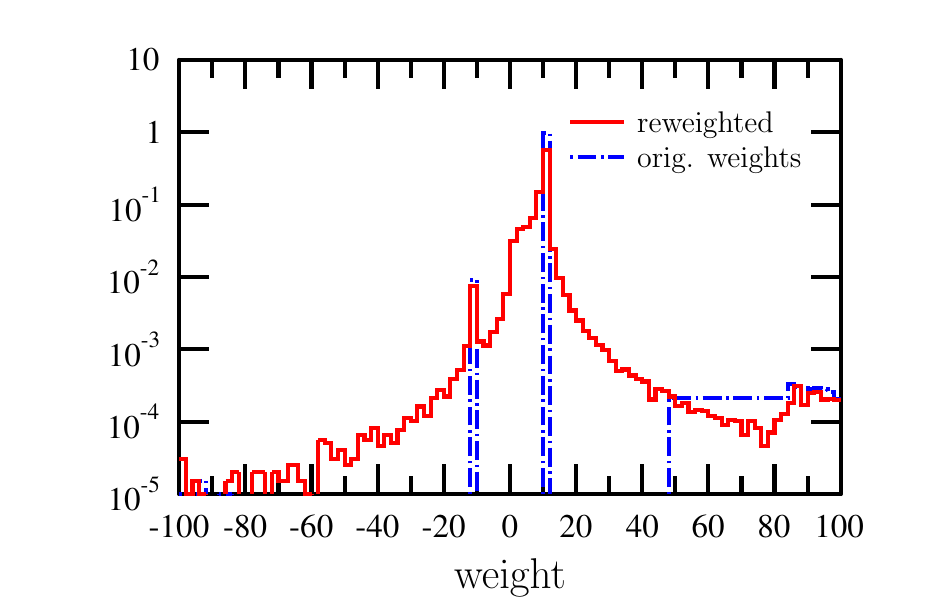}
\end{center}
\caption{Distribution of weights before and after reweighting.
}
\label{fig:weights}
\end{figure}

The majority of events generated by the \powhegbox\ have either
positive or negative, but in magnitude equal weights. As seen on
\fig{fig:weights}, the relative fraction of the negative weights is at
the percent level. (Due to the suppression in \eqn{eq:suppression} there
are also several large weight events starting at about 50 in
\fig{fig:weights}.) The drawback of the reweighting procedure is the
appearence of many different weights, leaving only about 90\,\% of the
events with equal positive weight.

\section{\label{sec:LHE} Predictions from the LHE's}

Using \powhel\ one can make predictions at four different stages in
the evolution of the final state: (i) at the parton level using NLO
accuracy, (ii) from the pre-showered POWHEG simulation (referred to
as LHE's), formally at the NLO accuracy, (iii) after
decay of the heavy particles, (iv) at the hadron level after full SMC.
While in an experimental analysis the last option is the most useful,
for examining the effect of the different stages in the evolution, and
also for checking purposes, it is useful to study the predictions at
intermediate stages. In the previous section we
mentioned that we made comparisons with existing predictions at stage (i)
and found agreement. We now show predictions at stage (ii), together with
the dependence on the choice of the scale $\mu_0$, and leave the
phenomenological analyses at the remaining stages to a more detailed
publication.

\begin{figure*}[t!]
\includegraphics[width=0.49\linewidth]{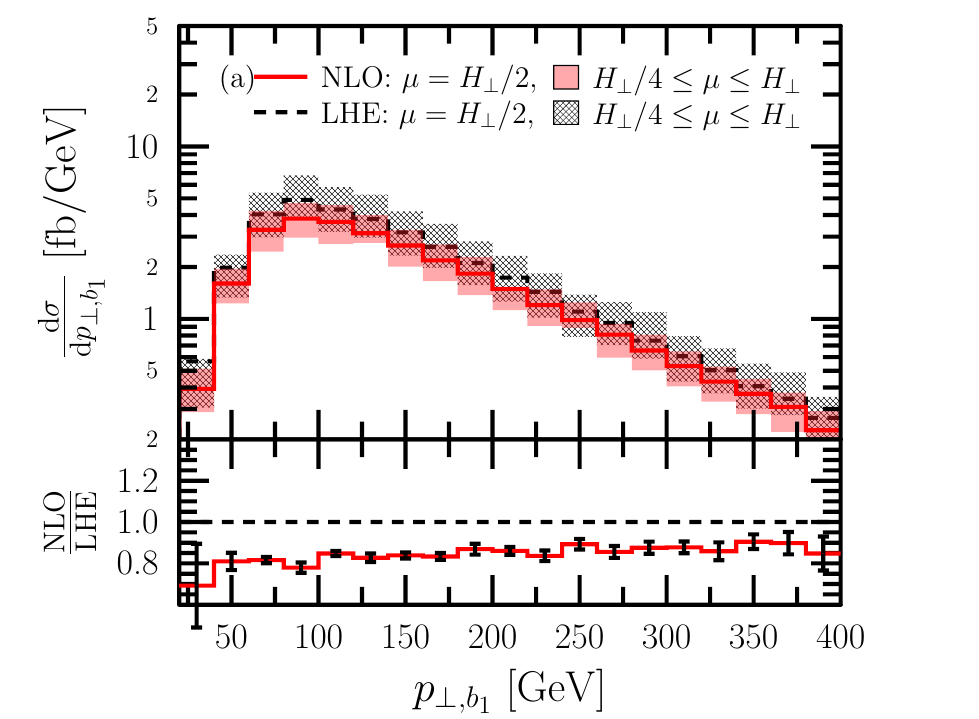}
\hfill
\includegraphics[width=0.49\linewidth]{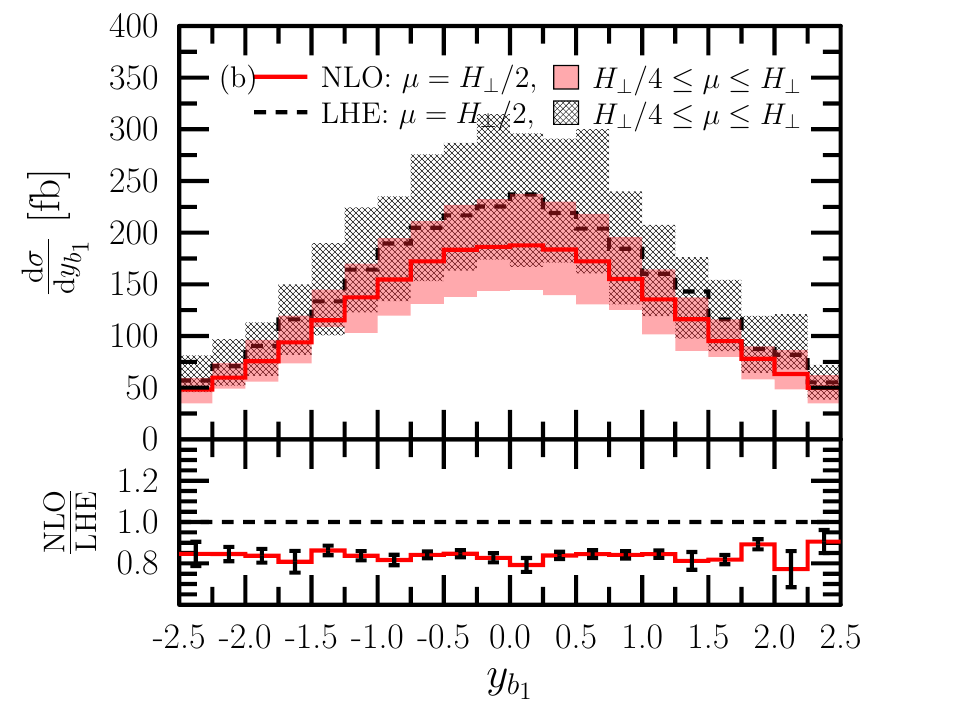}
\caption{Distribution of (a) transverse momentum (b) rapidity of the
hardest $b$-jet at the LHC at $\sqrt{s} = 14$\,TeV using 
\powhel. Distributions from LHE's are denoted LHE, while those at NLO
accuracy by NLO. The shaded bands correspond to cross sections obtained
with varying the scale around the default one in the range $[\mu_0/2,2 \mu_0]$. 
The lower panels show the NLO predictions normalized by the predictions
from LHE's, with errorbars representing the combined statistical
accuracy of the numerical integrations.
}
\label{fig:ptb1-lhe}
\end{figure*}

\begin{figure*}[t!]
\includegraphics[width=0.49\linewidth]{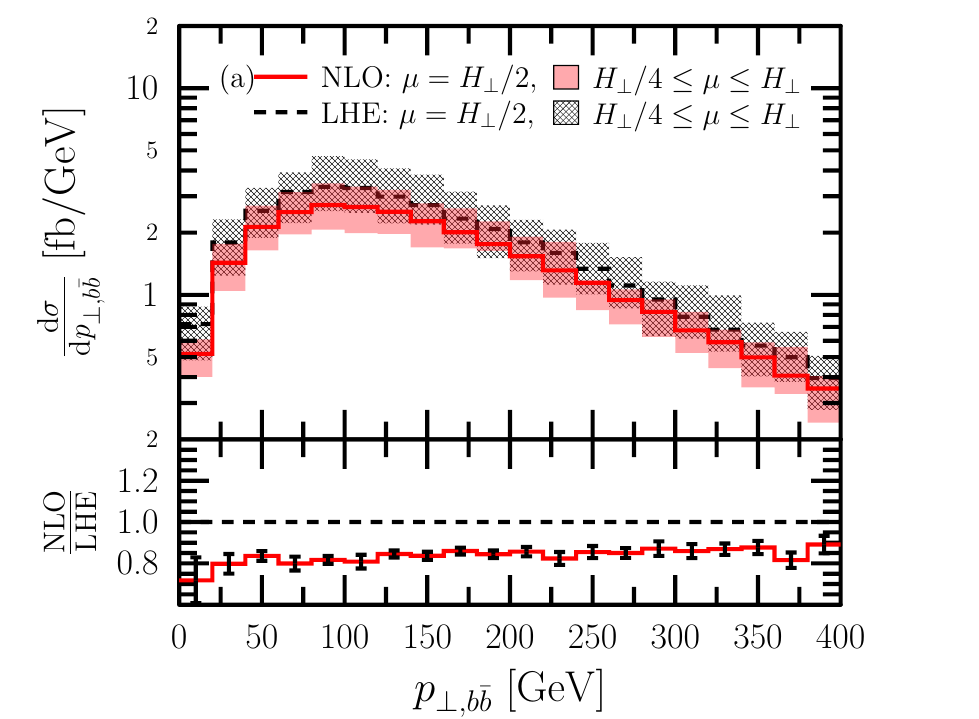}
\hfill
\includegraphics[width=0.49\linewidth]{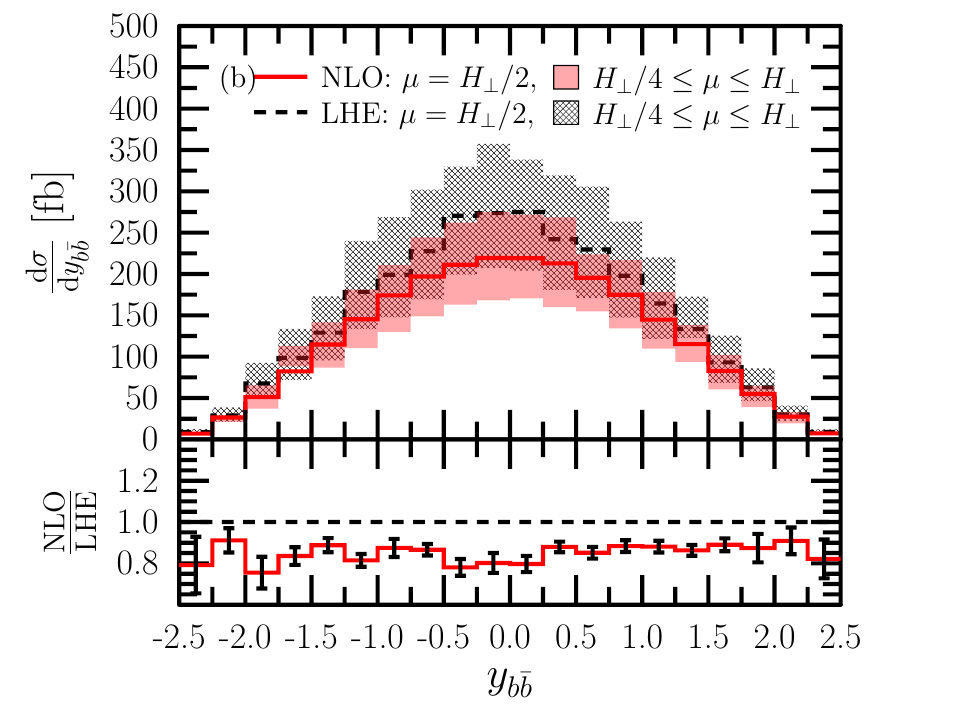}
\caption{Same as \fig{fig:ptb1-lhe} for the distribution of
(a) transverse momentum and (b) rapidity
of the $b\bar{b}$-jet pair.}
\label{fig:ptbb-lhe}
\end{figure*}

\begin{figure*}[t!]
\includegraphics[width=0.49\linewidth]{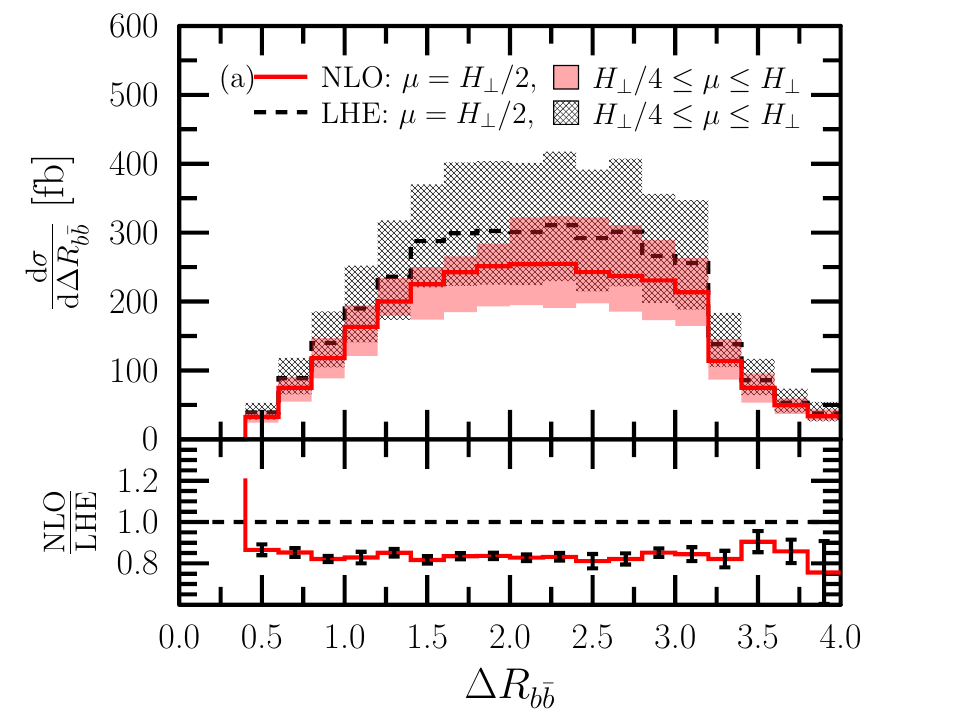}
\hfill
\includegraphics[width=0.49\linewidth]{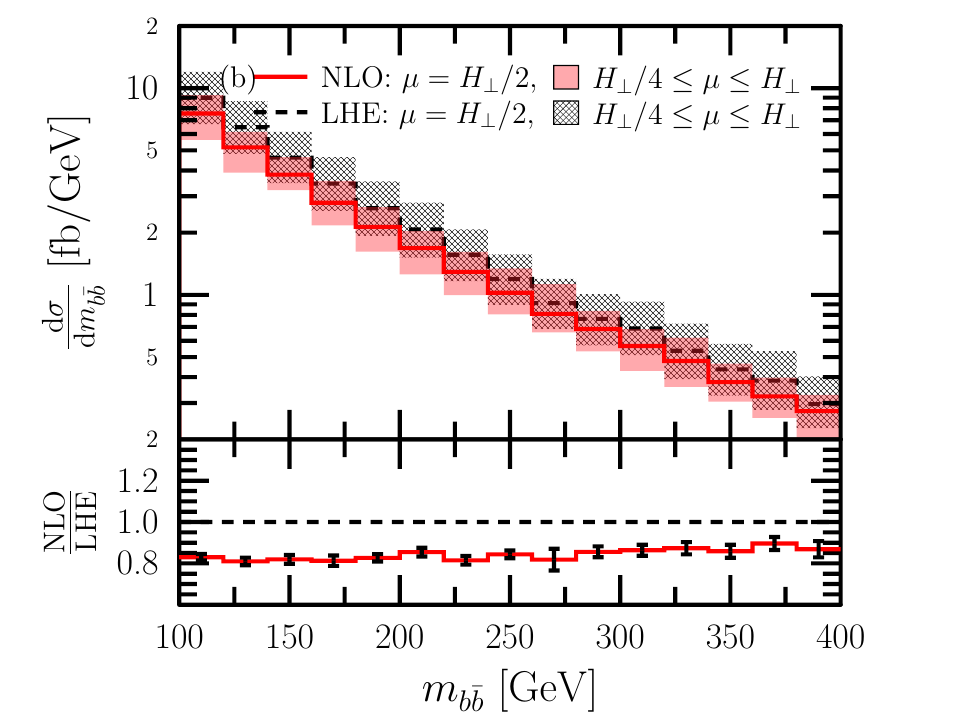}
\caption{Same as \fig{fig:ptb1-lhe} for the distribution of
(a) $\Delta R_{b \bar{b}}$ separation and (b) invariant mass
of the $b\bar{b}$-jet pair.}
\label{fig:mbb-lhe}
\end{figure*}

Our selection cuts are the same as in the NLO study above (setup 1 of
\Ref{Bredenstein:2010rs}).  In \figs{fig:ptb1-lhe}{fig:ptb2-lhe} we
show our predictions for kinematic distributions of the leading and
second hardest $b$-jet, respectively while in
\figs{fig:ptbb-lhe}{fig:mbb-lhe} we present our predictions for those
of the $b$-jet pair: transverse momentum and rapidity in \fig{fig:ptbb-lhe},
separation of the $b$-jets in rapidity--azimuthal angle plane
($\Delta R_{b\bar{b}}=\sqrt{\Delta y_{b\bar{b}}^2+\Delta \phi_{b\bar{b}}^2}$)
and their invariant mass in \fig{fig:mbb-lhe}.  The two bands
correspond to predictions obtained by varying the equal renormalization
and factorization scales in the range $[\mu_0/2,2\mu_0]$ around the
fixed default scale $\mu_0 = H_\bot/2$. If the renormalization
and factorization scales are varied independently in the same range,
the scale uncertainty is smaller.
\begin{figure*}[t!]
\includegraphics[width=0.49\linewidth]{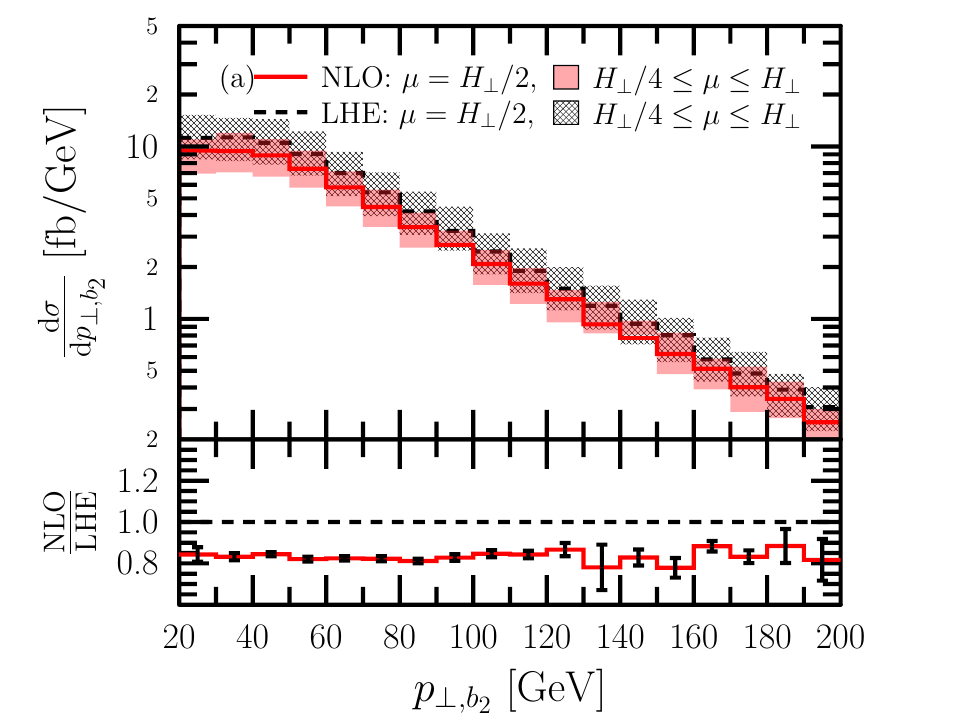}
\hfill
\includegraphics[width=0.49\linewidth]{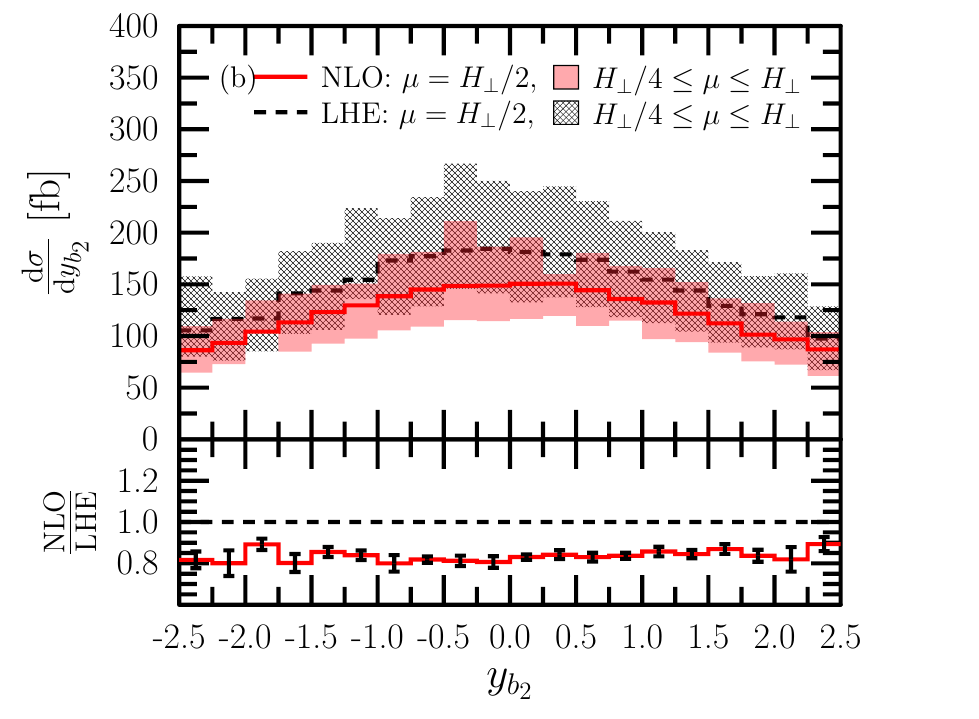}
\caption{Same as \fig{fig:ptb1-lhe} for the second hardest $b$-jet.
}
\label{fig:ptb2-lhe}
\end{figure*}

The distributions from the LHE's have the same shape as those at NLO
accuracy, but the normalization is larger by about 20\,\%. The two
predictions agree within the scale uncertainty of the NLO prediction,
which is the formal accuracy of the POWHEG method. Thus these LHE's can
be fed into SMC's for generating events at hadron level and perform
experimental analyses. 

\section{\label{sec:conclusions} Conclusions}

We have implemented the hadroproduction of a t\bt-pair in association
with a b\bb-pair, pp $\to$ t\bt\,b\bb, in the \powhel\ framework, which
can be used to generate LHE files. We discussed the technical subtleties
of such event simulations, in particular, issues related to the efficient
generation of the events for such a complex process: (i) the choice of
renormalization and factorization scales for generating the LHE's,
(ii) the implementation of higher precision arithmetics in computing
the loop amplitudes with sufficient precision, (iii) the advantage of
performing the event generation with fake virtual part and the method
of reweighting. 

The event files produced by \powhel, together with further detail and
results of our project, the corresponding version of the program
are available at http://grid.kfki.hu/twiki/bin/view/DbTheory.
We are confident that these LHE's are suitable input into SMC's to
produce distributions at the hadron level needed for experimental
analyses. Thus using those events one can make a detailed analysis of
t\bt$H$ signal and t\bt\,b\bb\ background predictions at the hadron level.

\subsection*{Acknowledgments}

We are grateful to M.V.~Garzelli, A.~van Hammeren, P.~Nason and
M.~Worek for useful discussions. This research was supported by
the Hungarian Scientific Research Fund grant K-101482,
the European Union and the European Social Fund through
LHCPhenoNet network PITN-GA-2010-264564,
and the
Supercomputer, the national virtual lab TAMOP-4.2.2.C-11/1/KONV-2012-0010
project.

\bibliographystyle{iopart-num}
\section*{References}
\bibliography{ttbb.bib}

\end{document}